\documentclass[onecolumn,9pt,  longbibliography]{asme2ej}
\usepackage{arydshln}
\usepackage{caption}
\newcommand{\AAs}{\AA\hspace{1ex}}  
\usepackage[hidelinks]{hyperref}
\usepackage{amsmath,amssymb,amsfonts, float, braket, mathrsfs, braket, bm}   
\usepackage{graphicx, subfig, epsfig, wrapfig}

\usepackage[sorting=none]{biblatex} 
\title{ 
Lossy Compression of Electron Diffraction Patterns for Ptychography via Change of Basis }
\author{Anton Gladyshev, Thomas C. Pekin, Marcel Schloz, Benedikt Haas, Johannes Müller and Christoph T. Koch*
 \affiliation{AG SEM, Department of Physics, Humboldt Universität zu Berlin \& IRIS Adlershof, Berlin, Germany\\Email: christoph.koch@hu-berlin.de\\ \today}}

\begin{document}
\maketitle

\begin{abstract}
\it{Ptychography is a computational imaging technique that has risen in popularity in the x-ray and electron microscopy communities in the past half decade. One of the reasons for this success is the development of new high performance electron detectors \cite{4d_stem} with increased dynamic range and readout speed, both of which are necessary for a successful application of this technique. Despite the advances made in computing power, processing the recorded data remains a challenging task, and the growth in data rate has made the size of the resulting datasets a bottleneck for the whole process. Here we present an investigation into lossy compression methods for electron diffraction patterns that retain the necessary information for ptychographic reconstructions, yet lead to a decrease in data set size by three or four orders of magnitude. We apply several compression methods to both simulated and experimental data - all with promising results.

}
\end{abstract}
\section{Introduction}

Used at synchrotron beamlines, in optical microscopes, or transmission electron microscopes (TEMs), ptychography recovers a two-dimensional or three-dimensional complex transmission function of an object from a four-dimensional dataset, comprised of a set of diffraction patterns that have been collected for many (often thousands) of (overlapping) beam positions on the sample \cite{4d_stem}. Through the use of a direct \cite{wigner,single_side_band}, or a more complex iterative \cite{Rodenburg04,PIE, ePIE,VandenBroek13,maximum_likelihood, lsq_ml, paper_rop} reconstruction algorithm, ptychography can provide a higher resolution than the conventional Rayleigh limit determined by a microscope's aperture \cite{Nellist95} and correct for microscope aberrations \cite{Nellist93,Sagawa16}. 

The foundations of ptychography were first described in a series of papers by Hoppe and coauthors in the '60s \cite{hoppe_1,  hoppe_2, hoppe_3, hoppe_4}, but a number of breakthroughs such as fast cameras, improved reconstruction algorithms and an increase of computational power had to be made before this technique started to provide results that overcame limitations of other imaging techniques, first in photon imaging (optical and X-ray) \cite{x-ray_ptychography, xray2, xray3}, and ultimately also in electron imaging \cite{Rodenburg08}. 

Nowadays, acquisition of 4D-STEM datasets \cite{4d_stem} is becoming routine, and one can choose between several ptychographic reconstruction algorithms \cite{Rodenburg08,4d_stem, PIE, wigner,single_side_band, paper_rop, Rodenburg04,ePIE,VandenBroek13,maximum_likelihood, lsq_ml} to process the data \cite{wigenr_exp, single_side_band, Putkunz12, epie_recon}. However, one of the main drawbacks of ptychography remains –– the ratio between the amount of information required to be recorded and the amount of the output information that can be reconstructed from it.  
The combination of partial spatial coherence of the illumination, detector size, desired resolution of the reconstruction, and the requirement to sample overlapping specimen areas by adjacent probe positions imposes a lower limit on the amount of acquired data. This data size can quickly reach the limits of the storage system used to record the data and thus limits the range of some experimental parameters crucial for the new trends in ptychography (e.g. multifocus ptychography \cite{multifocus} or ptychotomography \cite{ptychotomo}) such as the number of sequential scans that can be acquired or the size of field of view that can be scanned. 

In X-ray ptychography,  data compression \cite{xray_comp1, xray_comp2, xray_comp3} is mainly based on two concepts -- singular value decomposition that uses the fact that the diffraction intensities from neighboring and overlapping scan positions are typically correlated according to the ptychographic oversampling \cite{paper_rop} and on constrained pixel sums, where the reduction of the diffraction data is achieved by summation over given regions. The second concept is closely related to binning, a compression technique that binds together neighboring pixels, however the two approaches are not identical. The main advantage of the constrained pixel sums compression is that not the pixels in a given region, but only their sums are constrained, similar to Sudoku \cite{xray_comp1}. Another compression technique \cite{PtychoShelves, xray_comp3} is based on an assumption that the measured diffraction patterns are well described by the Poisson statistics. The measured intensities are rounded to 8bit unsigned integers. 

Binning, having its disadvantages mentioned above, is the most common compression strategy in electron ptychography \cite{paper_rop} and leaves plenty of room for novel compression techniques to arise, e.g. the usage of binary 4D-STEM data \cite{binary_4d_stem} with a linear reconstruction algorithm (single side band \cite{single_side_band}). Yet the specific geometry of electron diffraction patterns in the context of an iterative ptychographic reconstruction has not been fully explored. 

During the research we used ADORYM \cite{ADORYM}. It is one implementation of an iterative gradient-based ptychographic reconstruction algorithm capable of directly minimizing the discrepancy between experimental and model-predicted data expressed by specific choices of error metrics, optionally combined with various regularization terms (see also \cite{VandenBroek13,paper_rop}). It is written in Python and computes the necessary derivatives using pytorch \cite{pytorch} or autograd \cite{autograd}, two different automatic differentiation packages, that can be chosen as a backend. Initially developed for X-ray ptychography, ADORYM offers a generic forward model, includes multislice reconstruction, partial coherence of the probe and a flexible Python implementation. After replacing the dispersion law of photons by the one for electrons, we used ADORYM's default sparse multi-slice forward model for electron ptychography.

\section{Theory}

\subsection{Ptychography}
\label{section: ptychography}
A 2D ptychographic reconstruction algorithm retrieves amplitude and phase of the scattering object from diffraction patterns that are stored as a 4D-STEM data set in terms of probe positions $\bm{\rho}_p = \rho_{p,x}$, $\rho_{p,y}$ and spatial frequencies in the detector plane $k_{f,x}$ and $k_{f,y}$ (in the remainder of the paper, bold face variables indicate vectors). Here we will briefly explain main theoretical aspects of the reconstruction procedure.

The interaction of thin specimens with the incident probe is mathematically described as the real-space product of a two dimensional wavefunction representing the incoming probe $P^{(in)}(\bm{\rho}-\bm{\rho}_p)$ with a complex transmission function that represents the object $O(\bm{\rho})$. 
\begin{align}
P^{(out)}(\bm{\rho})=P^{(in)}(\bm{\rho}-\bm{\rho}_p)\cdot O(\bm{\rho}). \label{eq: interaction}
\end{align}

In real-space the two-dimensional transmission function $O(\bm{\rho})$ depends on the spatial coordinate $\bm{\rho}=(x,y)$ and  physically represents the modulation of amplitude and phase of an incident electron wave. The introduction of a relative coordinate $\bm{\rho}-\bm{\rho}_p$ is used to describe the position within the transmission function relative to different probe positions. Very thin specimens do not absorb electrons, but impose mainly a phase shift on the traversing probe wave function. For this reason ptychographic reconstructions typically aim to reconstruct the phase $\phi(\bm{\rho})$ of the object (transmission) function $O(\bm{\rho})$, which is directly proportional to the projected electrostatic potential $V_{proj}(\bm{\rho})$ of the specimen ($\phi(\bm{\rho}) = \sigma\cdot V_{proj}(\bm{\rho})$, where the interaction strength $\sigma$ depends on the incident beam energy). Absorption of traversing electrons by the 2D object function $O(\bm{\rho}) = \exp[j \phi(\bm{\rho})]$ can be included by allowing the potential and thus the phase to also have a (positive) imaginary part \cite{VandenBroek13}.

The diffraction patterns measured in a real scattering experiment are conventionally stored as 4D-STEM data sets and will be referred to hereinafter as $I^{(m)}_{\bm{\rho}_p, \bm{k}_f}$. Starting from an initial guess for the object function (e.g. $O(\bm{\rho}) = 1)$, one calculates the expected diffraction patterns $I^{(e)}_{\bm{\rho}_p, \bm{k}_f}$ as the squared modulus of the Fourier transform of the wave defined in Equation (\ref{eq: interaction}).

\begin{equation}
I^{(e)}_{\bm{\rho}_p, \bm{k}_f}= |\mathscr{F}_{\bm{\rho}\rightarrow \bm{k}_f} \left( P^{(out)}(\bm{\rho})\right)|^2
\end{equation}
\noindent
where $\mathscr{F}_{\bm{\rho}\rightarrow \bm{k}_f}$ denotes a Fourier transform with respect to $\bm{\rho}$ as a function of $\bm{k}_f$.

The discrepancy between measured and expected intensities is calculated at each scan position with a loss function $\mathscr{L}$. It can be formed by various error metrics (solely or in combination). For our calculations we used the squared $L_2$ norm of the wave magnitudes \cite{magnitude_loss}:

\begin{align}
    \mathscr{L}(\bm{\rho}_p)=\sum_{\bm{k}_f}
    \left ( \sqrt{I^{(e)}_{\bm{\rho}_p, \bm{k}_f}}-
    \sqrt{I^{(m)}_{\bm{\rho}_p, \bm{k}_f}}\right)^2.
    \label{eq: uncompressed_loss}
\end{align}

After the loss estimation the algorithm computes the derivatives with respect to the quantities to be optimized (the phase $\phi(\bm{\rho})$ and probe function $P^{(in)}(\bm{\rho})$) and updates them in a line search reducing the loss $\mathscr{L}$. The reconstruction iterates until the value of the loss function becomes sufficiently small.

\subsubsection{Multi-slice formalism}
To describe the interaction of the probe with thicker specimens, one has to account for multiple scattering events \cite{multislice}. In this case, the transmission function of the specimen cannot be treated as a two-dimensional quantity. The propagation of the beam is divided into intervals  $\left[z_{i}, z_{i+1}\right )$ and the projected object potential is broken into multiple two-dimensional slices $V_{proj}^{(i)}(\bm{\rho}) = \int_{z_{i}}^{z_{i+1}}V(\bm{\rho},z)dz$ along it resulting in object slices $O^{(i)}(\bm{\rho}) = \exp[j \sigma V_{proj}^{(i)}(\bm{\rho})]$. For a given
two-dimensional wave function of the incoming electron beam $P^{(in)}(\bm{\rho}-\bm{\rho}_p)$ entering the object at position $\bm{\rho}_p$, a 2D-wave $P^{out}$ exiting the object is computed by a sequence of propagation and interaction steps:

\begin{align}
    &P^{out}(\bm{\rho})= \left ( P^{in}(\bm{\rho}-\bm{\rho}_p)    \underbrace{  \cdot O^{i}(\bm{\rho}) }_{interaction} \right )   \underbrace{ \otimes_{\bm{\rho}} P_{Fr}(\bm{\rho}, z_{i+1}-z_{i})}_{propagation}      \label{eq: multislice_prop}\\
    &{P}_{Fr}(\bm{\rho}, d) =\mathscr{F}_{\bm{q} \rightarrow \bm{\rho}} \left\{ e^{-i\pi \lambda d |\bm{q}|^2}\right \}
\end{align} 

Here ${P}_{Fr}(\bm{\rho}, d)$ is a Fresnel propagator, $\otimes_{\bm{\rho}}$ denotes a real-space convolution, $\mathscr{F}$ represents a Fourier-transform and $\lambda$ is the de Broglie wavelength of the beam electrons. The subsequent calculations of the expected intensity and the loss function remain the same as described in subsection \ref{section: ptychography}.

\subsection{Compression methodologies}
For each probe position $\bm{\rho}_p$ diffraction patterns recorded by a detector with $N_x \times N_y$ pixels are conventionally stored as matrices, whose elements contain information about the intensities recorded at individual pixels. Mathematically, these matrices can be treated as elements of a $N_x \cdot N_y$  vector space of real numbers with the dot product $\langle .,. \rangle_{F}$ and norm $\|. \|_F$ computed by the Frobenius inner product. With two $N_x \times N_y$ matrices $\bm{A}$ and $\bm{B}$ this can be defined as follows:
\begin{align}
    \langle\bm{A}, \bm{B}\rangle_{F}=\sum_{ij}A^*_{ij}B_{ik}\\
    \|\bm{A} \|_F=\sqrt{ \langle\bm{A}, \bm{A}\rangle_{F}}
    \label{eqn:dotprod}
\end{align}

The traditional basis set for this vector space, hereinafter referred to as the "pixel basis", is composed of basis vectors representing single pixels. Using this basis set, one can describe any $N_x \times N_y$ matrix. However, the diffraction patterns recorded within a 4D-STEM experiment are quite similar in their intensity distributions. It should thus be possible to approximate them using a basis set having fewer elements than the original pixel basis. An ideal basis for compressing the data of a 4D-STEM experiment would satisfy the following requirements:

\begin{enumerate}
    \item Provide the best achievable resolution given the amount of data used. 
    \item Provide a good match between the original patterns and their compressed representations.
    \item Satisfy the previous two items with as few degrees of freedom as possible
\end{enumerate}

As previously, we assign a subscript $e$ to the expected patterns predicted by forward propagation and $m$ to the projected (measured) patterns stored in the 4D-STEM data set. The decomposition coefficients $c_{n}$ in a new basis were calculated for the square roots of both intensities as Frobenius inner products:
\begin{align}
    c_{n, \bm{\rho}_p}=\sum_{\bm{k}_f}\sqrt{I_{\bm{\rho}_p, \bm{k}_f}}\cdot M_{\bm{k}_f, n}
\end{align}

\begin{figure}
  \centering
  \includegraphics[width=\textwidth]{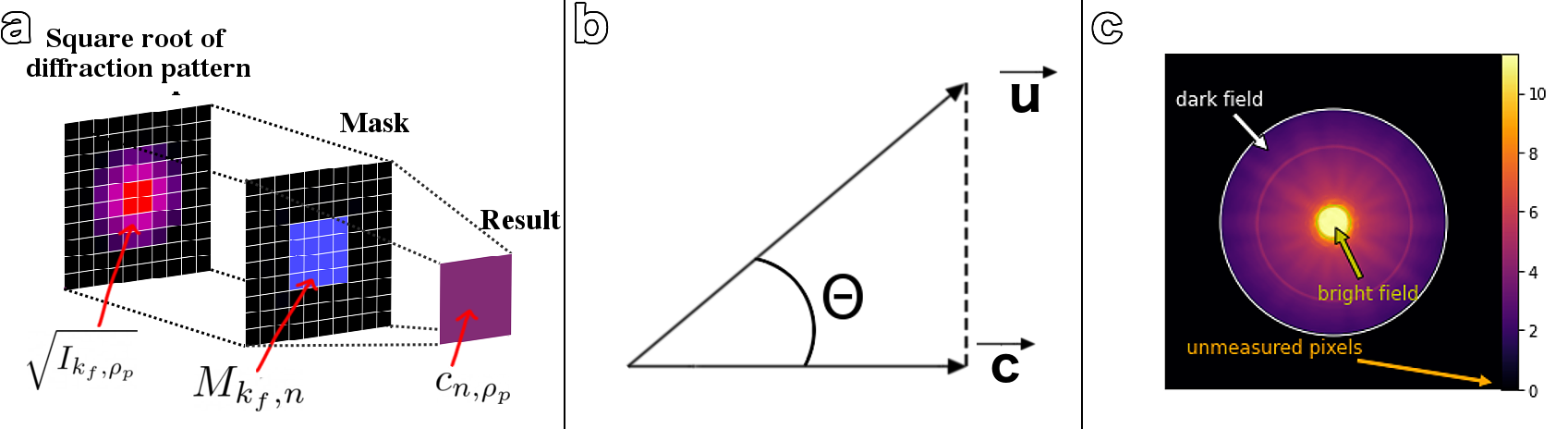}
  \caption{\textbf{a)} Schematic of the masking process. \textbf{b)} Illustration of the uncompressed ($\vec{u}$) and compressed ($\vec{c}$) representations of a vector. $\vec{c}$ and and $\vec{u}$ create an angle $\Theta$. A compression basis is considered suitable when it reduces this angle. \textbf{c)} A typical log-scaled electron diffraction pattern.
  }
  \label{fig:compression}
\end{figure}

Here $ M_{\bm{k}_f, n}$ is a pixel-basis representation of the n-th new basis vector. When  calculating the decomposition coefficients with a pytorch convolution layer within the ADORYM framework we interchange the terms "basis vector" and "mask". The panel \textbf{a} of Figure \ref{fig:compression} illustrates this masking process.

Fig.~\ref{fig:compression}c) shows a simulated electron diffraction pattern that can be separated in three non-overlapping regions: bright field, dark field and the outer regions containing unmeasured pixels. Because of this specific geometry, it is convenient to use two separate non-overlapping basis sets to describe the bright and dark field. The pixels in the unmeasured areas of the detector (in an actual experiment these would for example be shadowed by the HAADF detector) contain only zeros, so one does not have to store them.

\subsubsection{Loss function for compressed patterns}

When using the new basis set, the loss function $\mathscr{L}$ is computed not by comparing individual pixel values, but rather the decomposition coefficients:
\begin{align}
    \mathscr{L}(\bm{\rho}_p)=\sum_{n} \left (c^{(e)}_{n, \bm{\rho}_p}- c^{(m)}_{n, \bm{\rho}_p}\right)^2.
\end{align}

This loss function can be extended by various terms. Excluding the unmeasured pixels of the measured diffraction patterns from the dataset may cause divergence of the reconstruction. To prevent this a sum over the unmeasured (edge) pixels of the expected intensity can be added to the loss function:

\begin{align}
    \mathscr{L}_{edge}(\bm{\rho}_p)=\sum_{\bm{k}_f \in \text{ } edge}  I^{(e)}_{\bm{\rho}_p, \bm{k}_f} \label{eq: loss_edge}
\end{align}
which implicitly forces the values in those unmeasured pixels towards zero.

Panel \textbf{b} of Figure \ref{fig:compression} demonstrates a situation, where an uncompressed 2D-vector $\vec{u}$ is projected onto a compression axis, producing a compressed vector $\vec{c}$. In the context of our discussion, the squared length of  $\vec{u}$ can be interpreted as a total intensity of a diffraction pattern and $|\vec{c}|^2$ is the amount of intensity included by the new basis. We found it useful (see Appendix \ref{appendix: A}) to separately constrain the "floating parts" of the intensity in the bright and dark fields.

Knowing the total intensity of a pattern in the bright or dark field region of the diffraction pattern and the decomposition coefficients in the new basis, one can estimate the squared length of a "perpendicular" component $c^{(\perp)}$ not included into a new basis with the Pythagoras theorem:
\begin{align}
    \left(c^{(e/m)}_{\perp, \bm{\rho}_p}\right)^2=\sum_{\bm{k}_f} \left (I^{(e/m)}_{\bm{\rho}_p, \bm{k}_f}  \right ) - \sum_{n}\left (c^{(e/m)}_{n, \bm{\rho}_p}\right )^2 \label{eq: perp_comp}
\end{align}

In the Equation above $\bm{k}_f$ and $n$ correspond either to bright or dark field. For both bright and dark field detector areas the loss function contributions $\mathscr{L}_{\perp}(\bm{\rho}_p)$ were formulated as the $l^1$ norm of the difference between the expected and measured squared perpendicular components. A heavy weighting of this part of the loss function minimizes gross intensity differences in the diffraction patterns that were experimentally acquired and ones simulated in the ptychographic reconstruction.

Finally, we define a constraint for the probe.
The values of the phase-space wavefunction of the electron beam lying outside of the physical probe-forming aperture with a radius $k_a$ were constrained to be zero:
\begin{align}
    \mathscr{L}_{probe}=\sum_{|\bm{k}|>k_a} \left | \mathscr{F}_{\bm{\rho}\rightarrow \bm{k} }
    \left \{ P^{(in)} (\bm{\rho} )
    \right \} \right |^2. \label{eq: probe_constraint}
\end{align}

In addition to the loss-function contributions described above we considered two conventional regularization terms \cite{paper_rop}: total variation (TV) of the reconstructed object function and the $l^1$-norm of its phase. The total mismatch was calculated as follows:
\begin{align}
\mathscr{L}(\bm{\rho}_p)&=\underbrace{\sum_{n} \left | c^{(e)}_{n, \bm{\rho}_p}- c^{(m)}_{n, \bm{\rho}_p}\right|^2}_{\text{bright and dark}}+ 
\left|\left(c^{(e, bright)}_{\perp, \bm{\rho}_p}\right)^2- \left(c^{(m, bright)}_{\perp, \bm{\rho}_p} \right)^2 \right|+\left|\left(c^{(e, dark)}_{\perp, \bm{\rho}_p}\right)^2- \left(c^{(m, dark)}_{\perp, \bm{\rho}_p} \right)^2 \right|+ \nonumber\\
&+\mathscr{L}_{edge}(\bm{\rho}_p)+ \mathscr{L}_{probe}+ TV\left\{O(\bm{r})\right\}+ \sum_{\bm{r}}|O_{phase}(\bm{r})|. \label{eq: total_loss}
\end{align}
    
All parts of the loss function can be supplied with adjustable weights, which are dropped from the discussion in order to keep the expressions short. The main interest of this work was to investigate the behavior of the compressed loss-function, thence the weights of the conventional regularization terms were kept low. The parameters used in all further shown reconstructions are listed in Table \ref{table:parameters} in Appendix \ref{appendix: D}.

The form of the loss function defined in Eq. \ref{eq: total_loss} ensures that the predicted patterns match those measured along the directions specified by the basis, while the constraints try to make their total length in terms of the $L_2$ norm equal in separated areas of the detector plane. The impact of the constraints accounting for the total length is discussed in Appendix \ref{appendix: C}.

\subsubsection{Basis options}

In the following section we compare three orthonormal basis set systems: binning, Zernike polynomials, and a data-dependent basis set.

\begin{figure}[H]
  \centering
  \includegraphics[width=\textwidth]{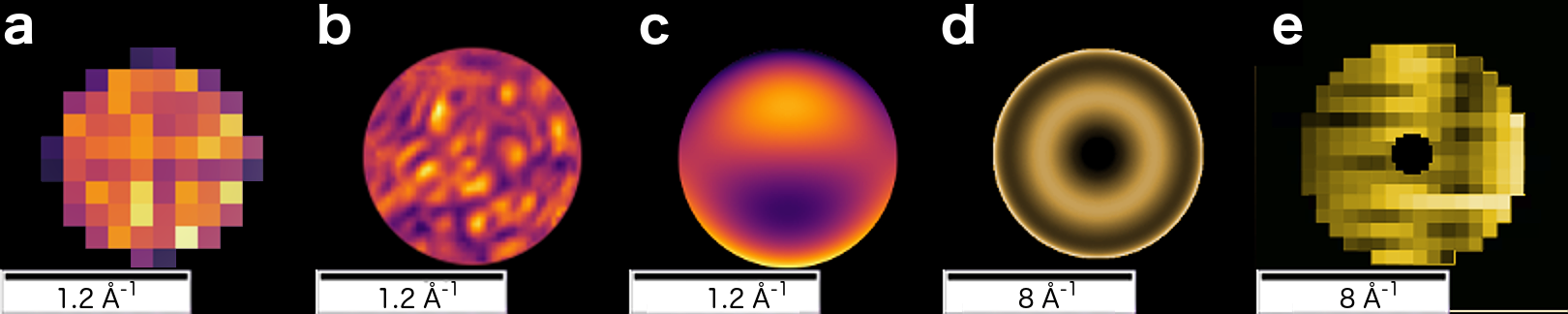}
  \caption{\textbf{a)} Bright field binning masks  (denoted with individual colors)
polynomial used for the description of the bright field.
  \textbf{b)} A mask generated from a prerecorded pattern.
  \textbf{c)} An example of a Zernike polynomial.
  \textbf{d)} A modified Zernike polynomial used for the description of the dark field. Note the change in scale. \textbf{e)} Binning masks used for the dark field description. The individual binning masks do not overlap, so all masks can be shown in a single picture. Even though the Zernike and diffraction pattern basis sets are orthonormal, the individual masks cover the same pixels, so it is not possible to show multiple masks in a single panel.
  }
  
  \label{fig:masks}
\end{figure}

Binning can be computed by using square-shaped non-overlapping masks that average the pattern value over an area (for an application of conventional binning to ptychography see, e.g. \cite{paper_rop}). The difference between conventional binning and the implementation here is a scaling factor. In the implementation described here an orthonormal basis is used in which the $l^2$ norm of each mask is equal to one. An example of bright field binning masks is shown in panels \textbf{a} and \textbf{e} of Figure \ref{fig:masks}.

An alternative option for the basis set are Zernike polynomials - an orthonormal set of polynomials within the unit circle. Two sets of Zernike polynomials were used -- one for the bright field  and the second one for the dark field. 
Zernike polynomials can be described using two indices $n$ and $m$.
\begin{align}
 & Z_{n}^{m}=N_n^m\cdot R_{n}^{m} (r) cos(\phi)\\
&Z_{n}^{-m}=N_n^m\cdot  R_{n}^{m} (r) sin(\phi)\\
&R^m_n(r) = \sum_{k=0}^{\tfrac{n-m}{2}} \frac{(-1)^k\,(n-k)!}{k!\left (\tfrac{n+m}{2}-k \right )! \left (\tfrac{n-m}{2}-k \right)!} \;r^{n-2k}
\end{align}
$N^m_n$ is a normalisation factor and $R_n^m$ is a radial part of the two dimensional Zernike polynomial $Z^m_n(r,\phi)$. The dark field polynomials have a hole in the bright field area and are thus doughnut-shaped. Because of that, Gram-Schmidt orthonormalization was used to make the modified Zernike dark field polynomials orthonormal to each other.

In principle one can use any arbitrary set of 2D orthonormal polynomials to describe a diffraction pattern. One nontrivial option is a basis set based on diffraction patterns acquired from the same sample, that the actual data is later recorded from. Due to the periodic structure of crystalline specimens, many of the recorded diffraction patterns are similar. Therefore, one can record diffraction patterns by a quick sampling across the sample (e.g. in an area adjacent to the area of interest or very coarsely in the same area) and use them  to create a basis for the subsequently recorded diffraction patterns comprising the actual data set. 

In the current work, diffraction patterns from random positions in the same area covered by the actual data set were used. To generate the compression basis, we used only the bright field region and applied a Gram-Schmidt algorithm to make the masks orthonormal. An example bright field mask generated in this way is presented in panel \textbf{b} of Fig. \ref{fig:masks}. This approach appears similar to a principal component decomposition \cite{PCA_1, PCA_2, PCA_3}, however it does not require the complete data set to be recorded before compression, since one selects the raw data randomly. Another benefit of our approach is that it is computationally much cheaper,  a new basis can be generated almost immediately and the compression can be applied directly during acquisition and before the data is stored to disk. Nevertheless, we have also tested a principal component compression approach, the results of which are presented in Appendix \ref{appendix: C}.

\subsection{Evaluation Metrics}
\subsubsection{Storage space}

In order to describe the relation between the storage space required for uncompressed and compressed 4D-STEM data sets, one can introduce a simple compression metric C. This is defined as the ratio between the degrees of freedom of the uncompressed pattern, i.e. the number of pixels, and the number of stored values per pattern:
\begin{align}
    C=\frac{\# \text{ of pixels}}{n_{bright}+n_{dark}+2}
    \label{eq: compression_metric}
\end{align}
\noindent
$n_{bright}$ and $n_{dark}$ describe the number of used bright and dark field masks, the term 2 corresponds to the perpendicular components $c^{(e)}_{\perp}$ one each for the bright and dark fields defined in the Equation \ref{eq: perp_comp}.

\subsubsection{Intact information metric}

In panel \textbf{b} of Fig. \ref{fig:compression} we showed the process of the dimension reduction from  2D-space to 1D-space. To describe how well the chosen basis system is able to match a given vector one can use the angle $\Theta$ presented in this Figure. Ideally, the vector to be compressed is an element of the space spanned by the basis, in which case the angle $\Theta$ is zero. In the worst case scenario all basis vectors are orthogonal to the chosen vector, and the angle $\Theta$ is equal to $\pi/2$. We calculated $\Theta$ separately for the bright and dark field areas of the data to estimate how well various basis systems describe the diffraction patterns. This metric is closely related to cosine similarity (in biology also known as Otsuka–Ochiai or Ochiai–Barkman coefficient \cite{cos_metric_1, cos_metric_2, cos_metric_3}), a more explicit discussion about this error metric can be found in Appendix \ref{appendix: A}.
\begin{align}
    \Theta_{bright/dark}(\bm{\rho}_p)=\arccos\left(\sqrt{\frac{ \sum_{n \in bright/dark} (c_{n, \bm{\rho}_p})^{2}}{
    \sum_{\bm{k}_f \in \text{ }bright/dark} I_{\bm{\rho}_p, \bm{k}_f}  
    }}\right).\label{eq: theta}
\end{align}
\subsubsection{Fourier shell correlation}

Fourier shell correlation (FSC) \cite{FSC_1, FSC_2} is a widely used metric for achieved resolution. It describes how well the phase-space representations of two real-space three dimensional data sets match each other. This is useful, for example, when the ground truth transmission function is known. For two object functions $O_1(\bm{r})$ and $O_2(\bm{r})$ the Fourier shell correlation (FSC) is calculated as follows:
\begin{equation}
    FSC(q)=\frac{\sum_{|\bm{k}|=q} \mathcal{F}^{*}_{\bm{r}\rightarrow \bm{k}} \left\{ O_1{(\bm{r})}\right\}\cdot \mathcal{F}_{\bm{r}\rightarrow \bm{k}} \left\{ O_2{(\bm{r})}\right\}  }{\sqrt{
    \sum_{|\bm{k}|=q} | \mathcal{F}_{\bm{r}\rightarrow \bm{k}} \left\{ O_1{(\bm{r})}\right\} |^2 \cdot  \sum_{|\bm{k}|=q} | \mathcal{F}_{\bm{r}\rightarrow \bm{k}} \left\{ O_2{(\bm{r})}\right\} |^2
    }}
\end{equation}

The Fourier-transformed images of atoms arranged in a perfect crystal are typically comprised only by very well defined (Bragg) peaks, with  mostly noise or zeros in between. Defects produce very weak signal between the Bragg peaks. In order to exclude the futile information, we evaluated the Fourier shell correlation only up to a cutoff-frequency of $2$ Å$^{-1}$. The value of the FSC is in general complex, we found it useful to evaluate the correlation using its real part.

\section{Simulated data}

\begin{figure}[H]
  \centering
  \includegraphics[width=\textwidth]{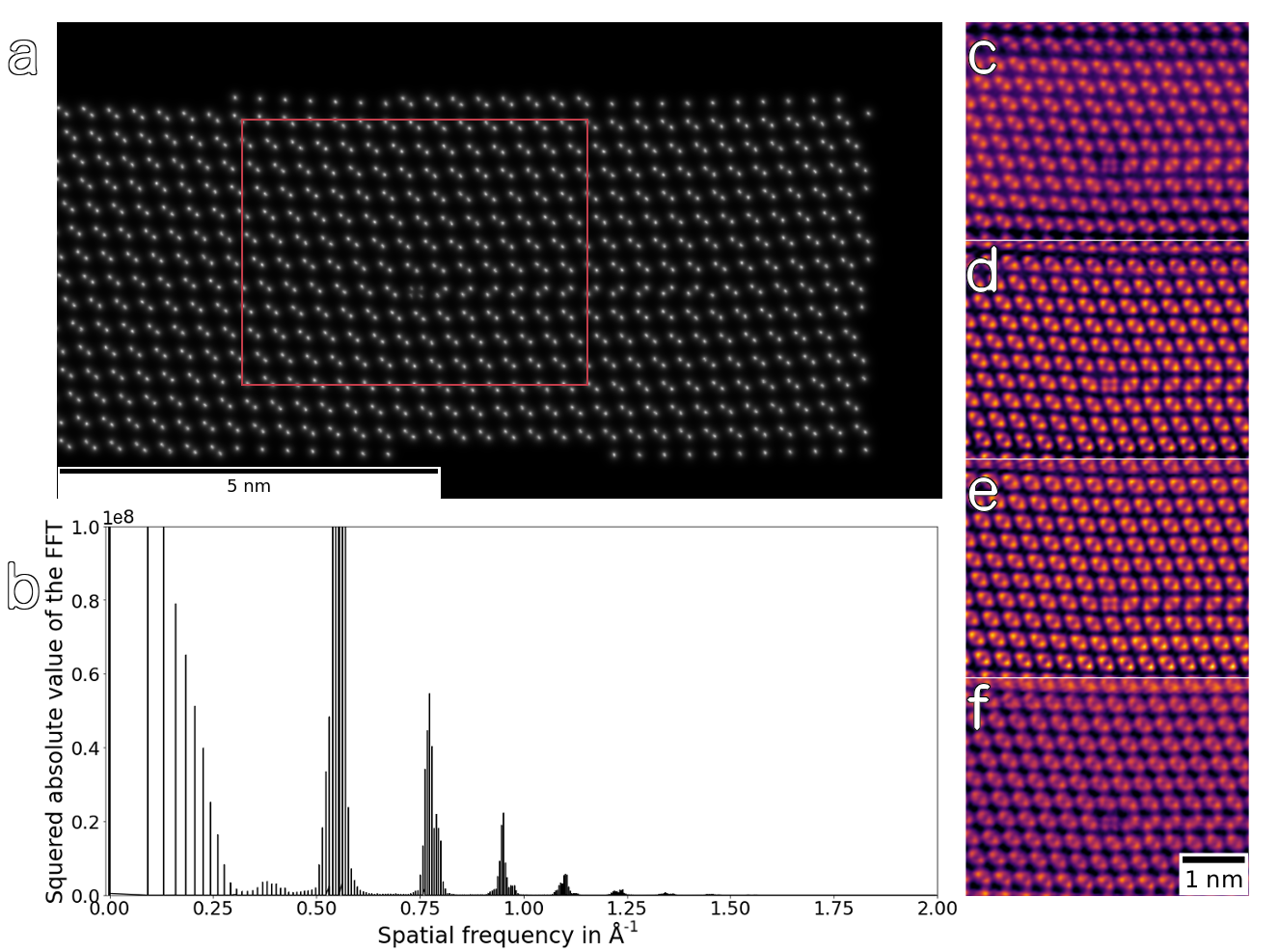}
  \caption{\textbf{a} Projected potential of the  92 \AAs thick Si dislocation model. Red box indicates the area presented in panels c)-f).  \textbf{b} Azimuthally integrated power spectrum  of the  projected potential at spatial frequencies $|\bm{q}|$.  \textbf{c}-\textbf{f} Four slices of the phase of the 3D object reconstructed from the uncompressed 4D-STEM data set. The parameters of the reconstruction are listed in Table \ref{table:parameters} in Appendix \ref{appendix: D}. }
    \label{fig:orig}  
\end{figure}

In order to be able to compare the reconstruction result with the ground truth, the compression was first tested using a simulated data set. This data set  was simulated using qstem \cite{Koch02}, including thermal diffuse scattering and partial spatial coherence. A 92 Å thick atomistic model of a silicon crystal containing a $60^{\circ}$ edge dislocation dissociated into a stacking fault bounded by a $30^{\circ}$ and $90^{\circ}$ partial dislocation \cite{Koch02,Spence06} was scanned by the probe with an accelerating voltage of 60 kV and with a step size of 1 \AAs along both, x- and y-direction. The probe was simulated for a convergence angle of 30 mrad using the following aberrations: defocus = 30 nm, astigmatism = 5 nm ($\phi_{A1} = 25^{\circ}$), coma = 500 nm ($\phi_{B2} = 80^{\circ}$), spherical aberration = -0.04 mm, chromatic aberration = 1.2 mm, energy spread = 0.3 eV, effective source size = 0.5 \AA. The simulated detector had 500x500 pixels covering scattering angles that corresponded to a real-space sampling of 0.07 Å in both dimensions. A gray-scale image presented in panel \textbf{a} of Fig. \ref{fig:orig} shows the projected potential of this model, the pseudo-color images in panels \textbf{c}-\textbf{f} show the phase maps that are proportional to the four slices of the object potential being reconstructed in this multi-slice ptychography reconstruction from the uncompressed data set. The upper and lower slice of the potential have weaker contrast than the two central slices, as one would expect from the limited resolution along the z-direction that corresponds to a beam convergence angle of 30 mrad at 60 kV. 

\begin{figure}
  \centering
  \includegraphics[width=\textwidth]{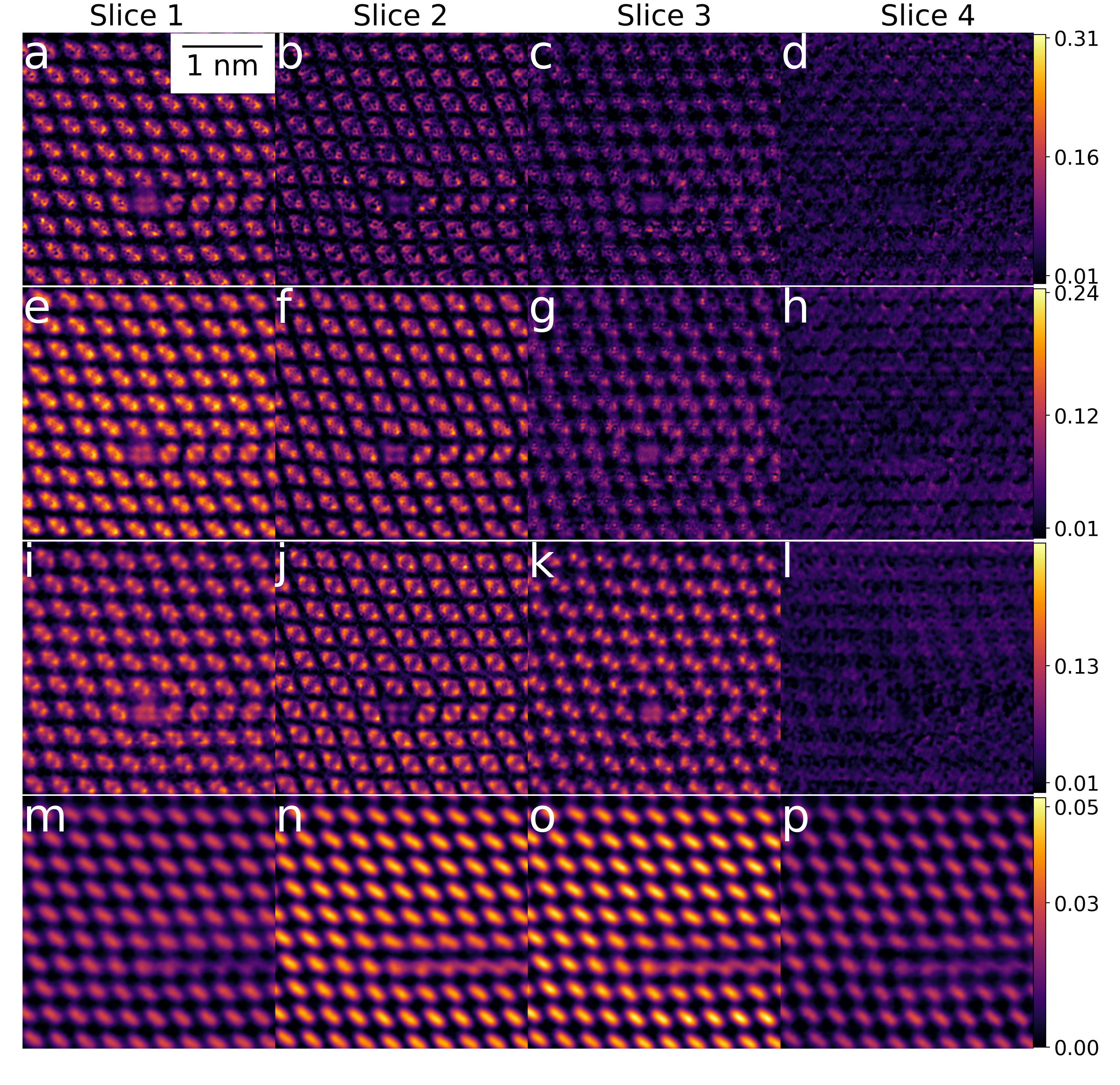}
  \caption{Multislice reconstructions of the phase of the object, reconstructed over four slices. The reconstruction in row \textbf{a)}-\textbf{d)} was done with the first 45 Zernike polynomials for the bright-field area of the detector and 1 dark field Zernike polynomial, in row \textbf{e)}-\textbf{h)} with the first 45 Zernike polynomials for the bright field area and 22 dark field Zernike polynomials for the dark-field area, in row \textbf{i)}-\textbf{l)} with 45 bright field and 1 dark field binning masks and in row \textbf{m)}-\textbf{p)} with 45 bright field and 172 dark field binning masks. The parameters of the reconstructions can be found in Table \ref{table:parameters} in Appendix \ref{appendix: D}.}
  \label{fig:dark_field}
\end{figure}

We found that the quality of the reconstruction improved with the number of bright field masks, independent of the decomposition basis. The object function converged with only one stored value accounting for the mean intensity in the case of all three dark-field basis sets (binning, Zernike polynomials and diffraction pattern basis), i.e. representing the total diffracted intensity. Storing more dark field values when using the Zernike basis only marginally improved the quality of the reconstruction. In Figure \ref{fig:dark_field}\textbf{a}-\textbf{h}, we compare two reconstructions that both used the first 45 Zernike bright field masks and 2 perpendicular coefficients defined by Equation (\ref{eq: perp_comp}), one each for bright and dark fields, but differed in the number of dark field masks. We used a single value (mean dark-field intensity) in \textbf{a}-\textbf{d} and the first 22 azimuthally invariant (with $m=0$) Zernike dark field masks in \textbf{e}-\textbf{h}. The reconstruction result improved slightly when including more dark-field masks. In Figs. \ref{fig:dark_field}\textbf{i}-\textbf{p} a similar comparison has been carried out for the case of compression by binning. Even though one would expect more dark field basis vectors to provide a more accurate description of patterns in terms of the metric defined in Equation \ref{eq: theta}, for compression by binning, the quality of the reconstruction worsens when using 172 dark-field masks (each being 22 px wide along both dimensions) versus a single dark-field mask (334 px wide along both dimensions). 

\begin{figure}[H]
  \centering
  \includegraphics[width=\textwidth]{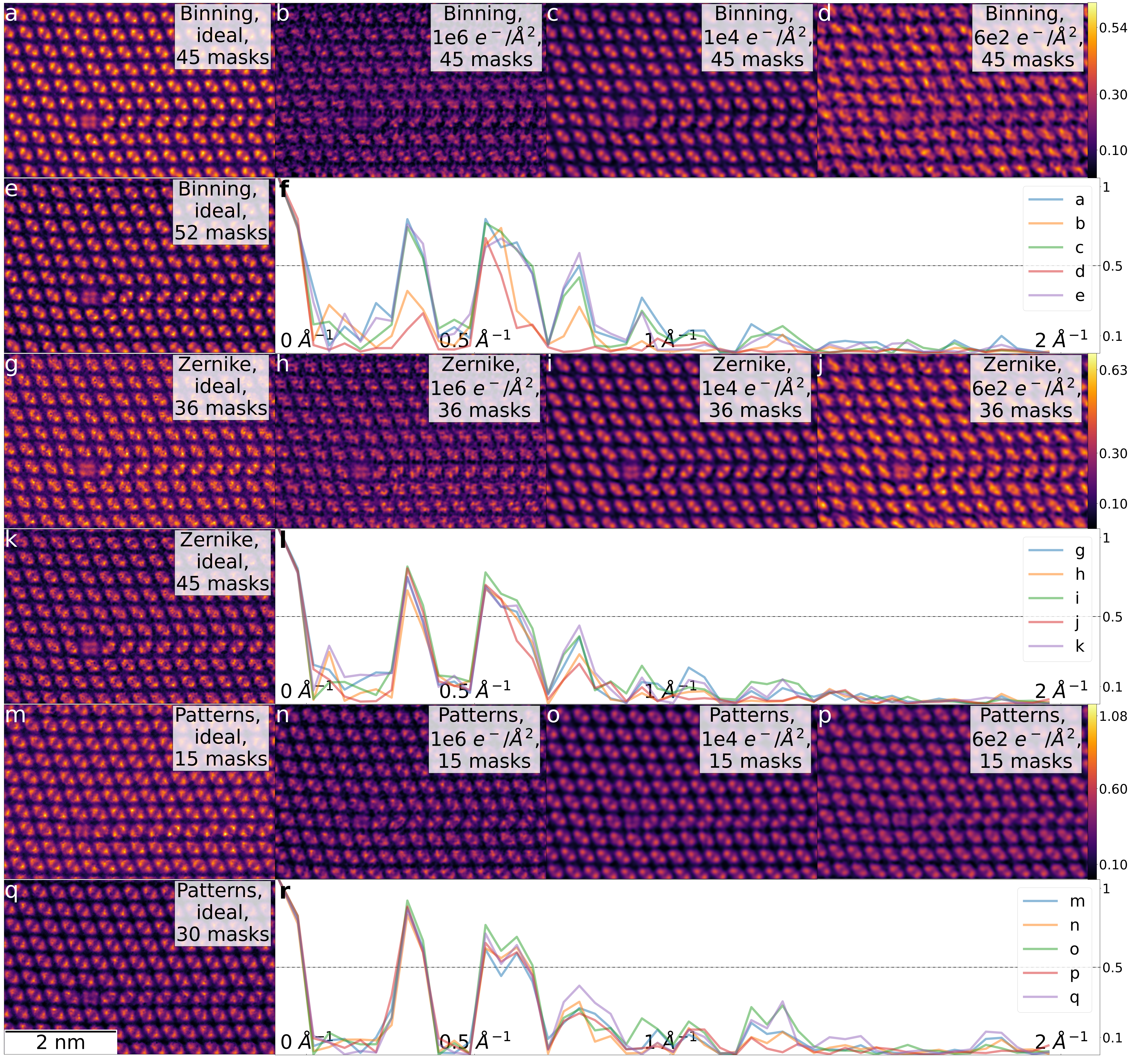}
  \caption{Summed phase along the beam direction of the reconstructed object using various compression basis sets and electron doses. Colorbar on the right is in radians. In all presented cases we used a single dark field mask and varied the number of bright field masks and the amount of noise. 
  The reconstructions \textbf{a)}, \textbf{e)}, \textbf{g)}, \textbf{k)}, \textbf{m)}, \textbf{q)} (first column) are done assuming ideal experimental situation, i.e. infinite dose and perfect coherence of the source. The reconstructions \textbf{b}-\textbf{d}, \textbf{h}-\textbf{j}, \textbf{n}-\textbf{p} are done assuming the partial spatial coherence and a finite dose, the value of which is given in the top right corner of the corresponding reconstruction. In \textbf{a)}-\textbf{d)} the bright field was described with 45 7 px wide binning masks, in \textbf{e)} with 52 6 px wide binning masks, in \textbf{g)}-\textbf{j)}  with first 36 Zernike polynomials, in \textbf{k)} with first 45 Zernike polynomials, in \textbf{m)}-\textbf{p)} with 15 masks generated from diffraction patterns generated before the main acquisition and in \textbf{q)} with 30 masks generated from diffraction patterns.The plots \textbf{f)}, \textbf{l)}, \textbf{r)} show the Fourier shell correlations of the reconstructed objects with ground truth, which was the potential used to simulate the 4D-STEM data. The parameters of the reconstructions are listed in Table \ref{table:parameters} in Appendix \ref{appendix: D}.
    } 
  
  \label{fig:result_comp}
\end{figure}
Fig. \ref{fig:dark_field} shows that the dimension reduction in the diffraction pattern makes the reconstruction slightly underdetermined. The shape of the reconstructed atoms varies from circular (e.g. panel \textbf{k}) to triangular (e.g. panel \textbf{j}), more noise is present (e.g. panel \textbf{b}) and/or the overall resolution of the reconstruction is relatively low in some of the cases (panels \textbf{m} - \textbf{p}). 

Another important aspect is that the probe $P^{in}(\bm{\rho})$, which was optimized simultaneously with the object during the reconstructions, converges to different defocus values depending on the basis set. Thus, some reconstructed slices in Fig.~\ref{fig:dark_field} (e.g. panels \textbf{d}, \textbf{h} and \textbf{l}) of the object are empty and the object appears to end.

After finding the optimal approach for the dark field compression, we investigated the limits of compression in terms of the minimal number of bright field masks and the lowest electron dose. Convergence of the reconstruction could be achieved by using only the first 36 Zernike masks, 45  binning masks (each 7 px wide and high) or 15 diffraction pattern basis masks to describe the bright-field region of the detector. A single coefficient describing the mean intensity in the dark-field area was used in all 3 cases. It corresponds to the signal one would collect using an ADF detector with a large outer radius and an inner radius being slightly larger than the semi-convergence angle of the illuminating probe.

With these limits discovered, the effect of a finite electron dose were explored.
We considered an effective source of size 0.5 Å and used a Poisson noise model to emulate the finite electron doses indicated for each of the panels in Fig. \ref{fig:result_comp}. Aside from some fluctuations in apparent reconstruction quality induced by the randomness of the applied noise, the results (phases summed over all 4 slices) shown in Figure \ref{fig:result_comp} are consistently better for the diffraction pattern basis (row 5). This is reflected in both the visual appearance of the phase maps across the noise levels as well as the FSC plots, especially for spatial frequencies corresponding to Bragg peaks. Fig. \ref{fig:orig}.\textbf{b} shows that the Fourier coefficients between the Bragg peaks are only very weakly represented in the original potential, which explains their poor representation in the FSC of in all three cases of compression basis. However, the FSC of spatial frequencies around 0.3 and 0.6 \AA$^{-1}$ are the highest for the diffraction pattern basis, even though only 15 coefficients have been used, in comparison to 36 in the case of Zernike polynomials and 45 in case of Binning. It is, however, noticeable that the diffuse signal between the Bragg peaks is represented weaker in the case of the diffraction pattern basis than in both other cases. Besides the lower number of coefficients, this may also be due to the random choice of diffraction patterns, which makes it likely that the perfect crystal information is more faithfully represented by this basis than all the defect configurations possible. 

\section{Experimental data}

With the various compression bases' suitability for ptychography confirmed in simulation, we turn to experimental data.
Although the authors of \cite{magnitude_loss} have shown that using the magnitude of a wave, i.e. the square root of its intensity, is more beneficial for the calculation of the loss than the intensity itself, the calculation of the square roots during on-the-fly compression may dramatically affect the acquisition time, depending on the recorded frame size and computational resources. Moreover, there already exist a number of tools that allow the use of the compression approach described above directly on the intensity without any adjustments.

Several software packages implement the computation of the dot product of a diffraction pattern with any mask as described by expression (\ref{eqn:dotprod}) very efficiently. As an example, the software driving the Dectris ELA detector attached to the Nion HERMES microscope used for acquiring the experimental data in this section applies this operation to the data `on the fly' at frame rates up to 100,000 frames per second \cite{Plotkin-Swing22}. From the perspective of the microscope operator, the diffraction patterns are thus transferred from the microscope already decomposed into the chosen basis set, obviating the need to store hundreds of gigabytes of data which must then later be decomposed. 

\begin{figure}[H]
  \centering
  \includegraphics[width=\textwidth]{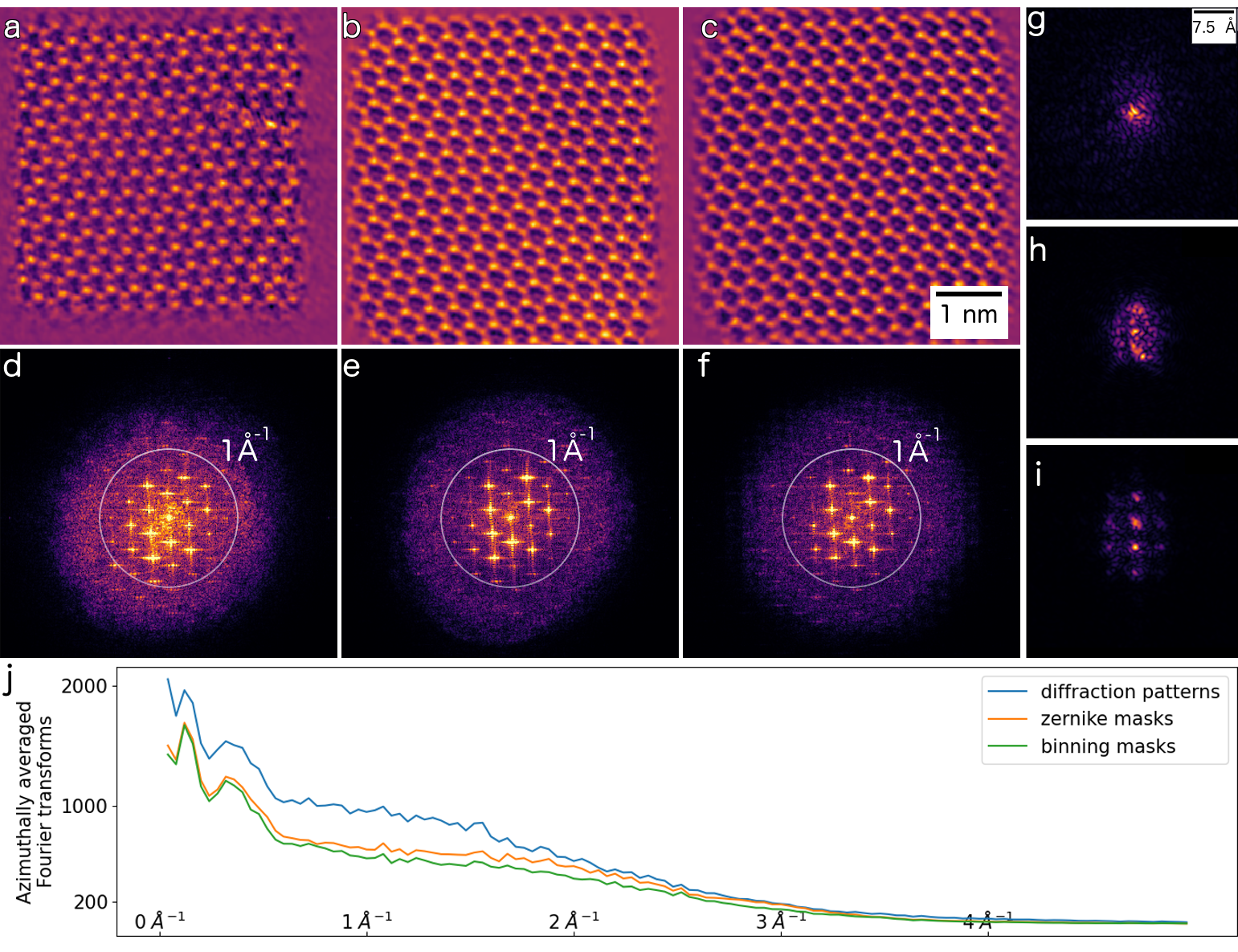}
  \caption{\textbf{a}-\textbf{c} Reconstructed phases of a complex transmission function of a MoS$_2$ monolayer from compressed 4D-STEM datasets. $128\times 128$ px wide diffraction patterns were compressed using a single dark field mask accounting for the mean dark-field intensity and multiple bright field mask configurations: \textbf{a)} 15 bright field masks generated from diffraction patterns, \textbf{b)} the first 36 Zernike polynomials, and \textbf{c)} 45 bright field binning masks (each $8 \times 8$ px in size). The diffractograms of these reconstructions shown in \textbf{d}-\textbf{f} are the log-scaled amplitudes of the Fourier transforms of the complex transmission functions presented in panels \textbf{a}-\textbf{c}, respectively. The white circles indicate spatial frequencies of 1 \AA$^{-1}$.  In \textbf{g}-\textbf{i} we show the magnitudes of the reconstructed probe functions that correspond to the reconstructions \textbf{a}-\textbf{c}, respectively. The plot \textbf{f)} shows azimuthal averages of the Fourier transformed object functions. The parameters of the reconstruction are listed in Table \ref{table:parameters} in Appendix \ref{appendix: D}.}
  \label{fig:exp_results}
\end{figure}

A monolayer MoS$_2$ was scanned at an accelerating voltage of $60$ kV with a convergence semi-angle of $33$ mrad with a step size of $0.2$ \AA.
$128 \times 128$ px wide diffraction patterns were compressed using a single dark field mask accounting for the mean dark-field intensity and the following three different compression bases for the bright-field region: 1) 15 masks generated from recorded diffraction patterns; 2) the first 36 Zernike polynomials and 3) 45 binning masks $8 \times 8$ pixels in size. Subsequently the dark field mask was padded so that the pixel width in real space was $0.138$ Å. To reduce the storage space we decided to neglect the perpendicular components $c_{\perp, \bm{\rho}_p}$ defined in the Equation \ref{eq: perp_comp}. In Figure \ref{fig:exp_results} we show the results of the reconstructions. 

The probe functions that have been reconstructed along with the object transmission functions show very clearly that only in the case of the diffraction pattern basis the probe does not seem to contain any object information, although this data was compressed the most. In both, the Zernike and binning compression the probe contains side lobes that correspond to inter-atomic distances of the MoS$_2$-lattice. This is particularly strong in the case of compression by binning. 

In addition to the probe functions, also the reconstructed object transmission functions itself is more faithfully reconstructed in the case of the diffraction pattern basis. The reconstruction shown in Fig. \ref{fig:exp_results}a (and its diffractogram in Fig. \ref{fig:exp_results}d) contains more diffuse scattering as well as defects,  which is physically more reasonable, since MoS$_2$ not sandwiched in hBN is typically covered by contamination \cite{Shao22}.

\section{Conclusion}
We have shown for the case of atomic-resolution ptychography in the STEM, data can be compressed by a factor of 5,000 to 15,000, while still achieving a resolution comparable to the uncompressed reconstruction. Three different compression approaches have been compared. For both simulated and experimental data the compression approach based on orthogonal masks produced from randomly recorded diffraction patterns seems to preserve a sufficient amount of detailed real-space  information to allow for a faithful reconstruction even in the case of reconstructing multiple object slices.  The estimation of the amount of ``intact" information is discussed in Appendix \ref{appendix: A}. 

For the presented reconstructions, it was sufficient to represent the dark-field intensity of the diffraction patterns by a single value accounting for the mean intensity, although a better knowledge of the dark field provides a slightly better result (see in Appendix \ref{appendix: B}), albeit at severely increased storage needs. The loss function used contained only two conventional regularization terms: total variation (TV) of the reconstructed object function and the $l^1$-norm of its phase, both with relatively small weights (see in Appendix \ref{appendix: D}). It has been shown that regularization can prevent artifacts due to underdeterminedness of the solution (see, e.g. \cite{paper_rop}), however, in Appendix \ref{appendix: C} we show that the $l^1$ constraint does not help to prevent the artifacts appearing in the reconstructed object. This fact suggests the need to develop new, more complex concepts of regularization that would specifically account for the compressed nature of the input data and potentially improve the visual appearance of the reconstructions. The investigation of regularization constraints, as well as the evaluation of the strength of improvement  of including more bright-field vs. more dark-field compression coefficients on the reconstruction quality are reserved for subsequent studies.

The presented tests with experimental and simulated data show that our compression algorithm is suitable for real experiments and works also for reconstructing objects under strong multiple scattering, realistic electron doses and partial spatial and temporal coherence. The new basis sets proposed here -- Zernike polynomials and a diffraction pattern basis require less storage capacity than the conventional compression by binning, allowing for more patterns to be acquired during an experiment and a more efficient use of data storage. 

\section*{Acknowledgements}
The authors acknowledge financial support by the Volkswagen Foundation (Initiative: "Experiment!", Project "Beyond mechanical stiffness") and by the Deutsche Forschungsgemeinschaft (DFG) in project nr. 182087777 (CRC951) and project nr. 414984028 (CRC1404).

\printbibliography

\newpage
\appendix
\section{Measurement of basis quality}
\label{appendix: A}

Here we show the relationship between the angle $\Theta$ defined in the Equation \ref{eq: theta} and the numbers of basis vectors in the bright and dark fields for the simulated 4D-STEM dataset.

\begin{figure}[ht!]
  \centering
  \includegraphics[width=\textwidth]{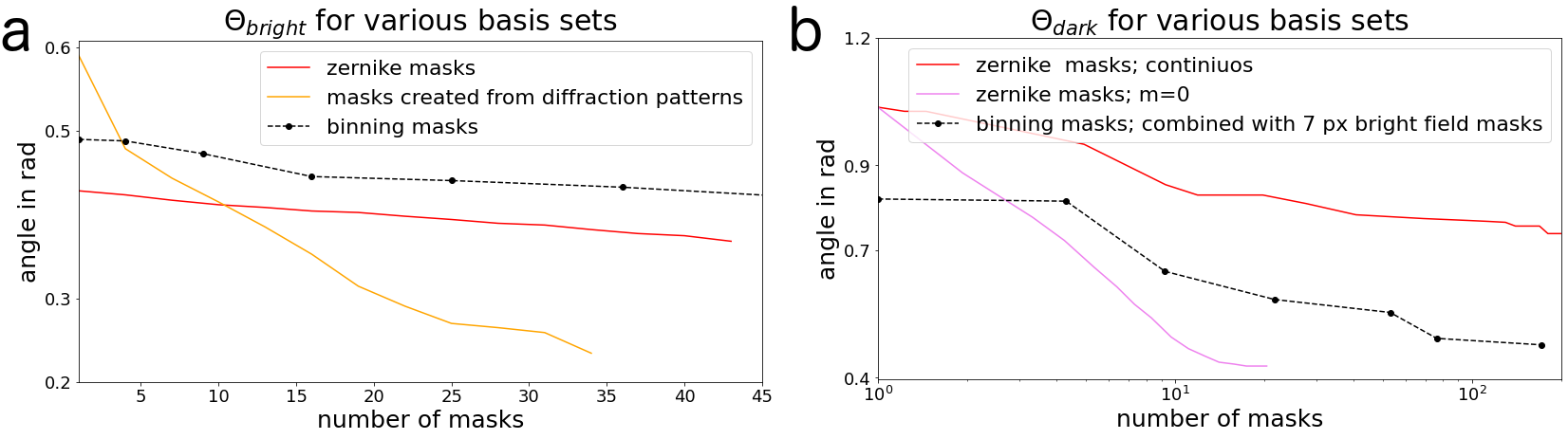}
  \caption{Angle $\Theta$ described in Equation \ref{eq: theta} averaged over multiple scan positions for bright- and dark-field areas of the diffraction pattern and various basis systems.  \textbf{a)} $\Theta_{bright}$ (mean angle enclosed by the bright field of a diffraction pattern and its projection on the bright field basis) for Zernike polynomials (red line), masks generated from randomly selected diffraction patterns (orange line) and binning masks (black dashed line). \textbf{b)} $\Theta_{dark}$ (mean angle enclosed by the dark field of a diffraction pattern and its projection on the dark field basis) for Zernike continuous  (all values of $m$) indexing (red line), azimuthally symmetric ($m=0$) Zernike polynomials (violet line)  and dark field binning masks whose central area has been cropped so that they can be combined with 7px wide bright field binning masks (black dashed line). 
  }
  \label{fig:amounts}
\end{figure}
The plot presented in panel \textbf{a)} of Figure \ref{fig:amounts} shows that a single mask generated from randomly selected diffraction patterns on average describes the diffraction patterns worse then the Zernike or binning masks. However, as the number of masks increases, the angle $\Theta_{bright}$ decreases much faster for this basis set and only 10-15 masks generated from randomly selected diffraction patterns perform noticeably better then a similar number of Zernike or binning masks.

In contrast to Zernike masks that perfectly matched to the geometry of the diffraction patterns, 7 px wide bright field binning masks extend slightly out of the true bright field area and partially cover the transition areas between the bright and high angle annular dark field (HAADF). These transition pixels carry values that are higher than the average count value in the rest of the dark field. Thus, as the panel \textbf{b)} of Figure \ref{fig:amounts} shows, a single binning mask that ignores these pixels describes the dark field better then the the first Zernike dark field mask.

It is noteworthy that the angle $\Theta$ is not a universal tool. To determine whether the reconstruction will converge, one has to consider the effects caused by partial spatial coherence, finite dose, and combine it with a total over-determination ratio in the pixel-basis, derived in \cite{paper_rop}.

\newpage
\section{Anomalous effects in regions not covered by the basis set}\label{appendix: B}

\begin{wrapfigure}{r}{0.5\textwidth}
\begin{center}   
    \includegraphics[width=0.5\textwidth]{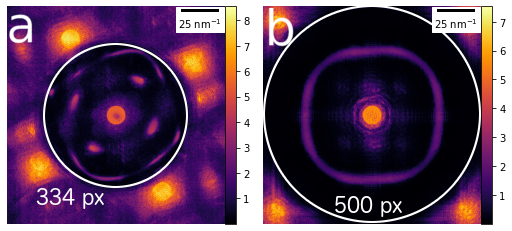}
\end{center}

  \caption{Magnitudes of the reconstructed probes in phase space when using  Zernike masks covering a circle with a diameter \textbf{a)} 334px and  \textbf{b)} 500 px. In both cases, the probe intensity in the uncovered area grows uncontrollably. }
  \label{fig:probes_bad}
\end{wrapfigure}

When performing a dimensional reduction of a vector space, only some part of the  initial information is saved. Even though the measured patterns might be well described with the chosen basis, the algorithm has no control over its predictions in this directions.  

This problem can be demonstrated on an example with uncovered areas. We attempted to use Zernike polynomials which only covered  a part of the detector, i.e. not including the edges of the detector, while simultaneously optimising probe and objects functions. In Figure \ref{fig:probes_bad}.\textbf{a}, we show a reconstructed probe function in phase-space. The bright field polynomials covered the pixels up to 44 px away from the center, the dark field polynomials covered the pixels with distances between  44 px and 334 px, and the pixels that are 335 px or further away from the center were completely ignored. During optimization, the probe collected artifacts in the uncovered/ignored areas. 

We discovered the same behaviour of the reconstruction also at different radii of the dark field polynomials. Figure \ref{fig:probes_bad}.\textbf{b} shows a Fourier transform of a reconstructed probe function. The  dark field polynomials used covered the pixels up to 500 px away from the center. In this case, the probe also collected artifacts in the uncovered areas. 

\begin{figure}[ht!]
  \centering
  \includegraphics[width=\textwidth]{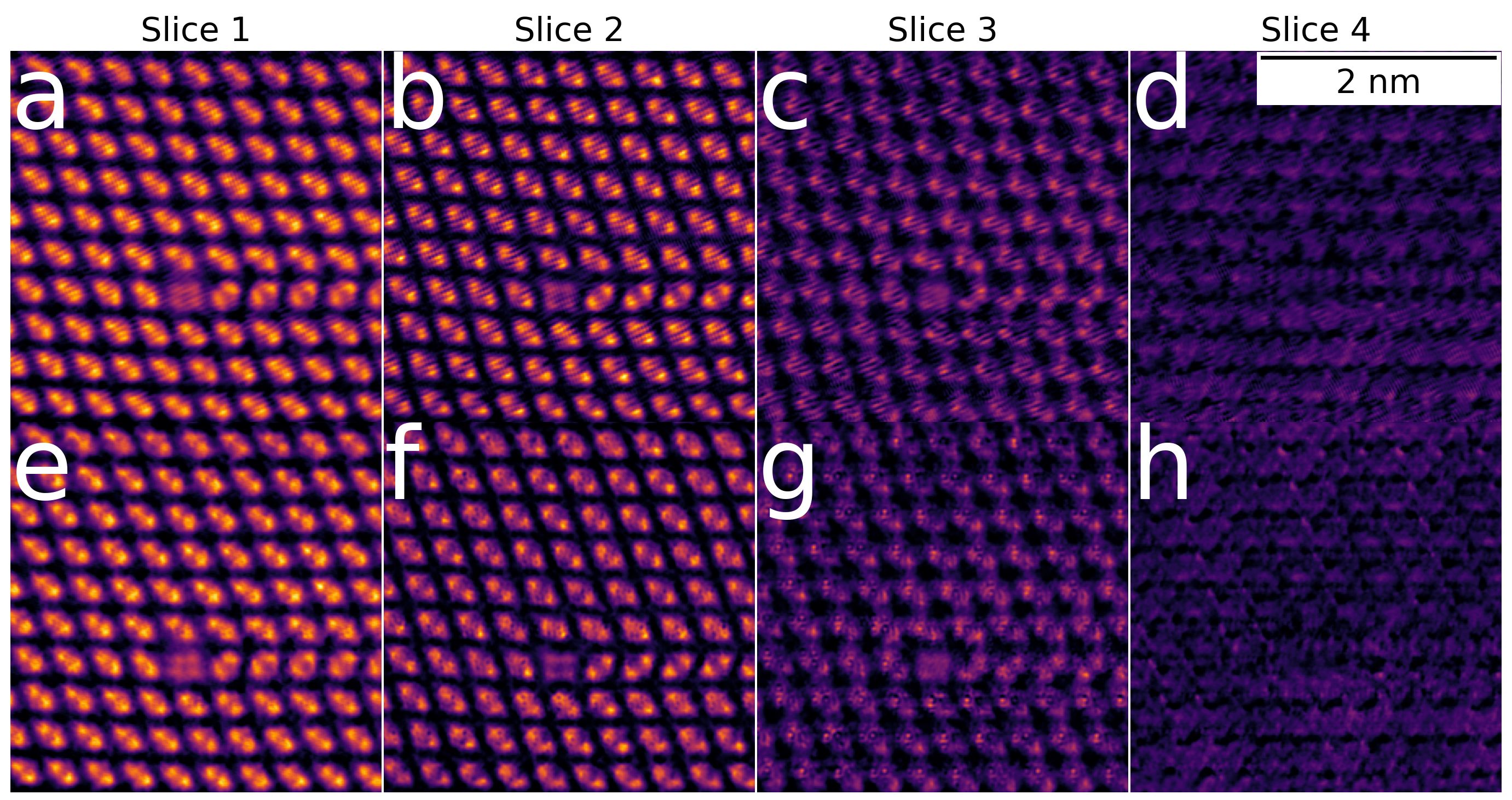}
  \caption{Two multislice reconstructions of the phase of the object, reconstructed over four slices that are positioned along the beams propagation direction at 0 Å, 30.7 Å, 61.4 Å and 92 Å, respectively. Both reconstructions \textbf{a)}-\textbf{d)} and \textbf{e)}-\textbf{h)} are done using first 45 Zernike bright field masks and first 22 angular-symmetric dark field Zernike masks. The reconstruction \textbf{a)}-\textbf{d)} was performed without constraint defined in the Equation (\ref{eq: perp_comp}).
  The reconstruction \textbf{e)}-\textbf{h)} was performed without it. It is apparent that in \textbf{a)}-\textbf{d)} the object function collected artifacts in form of stripes, while \textbf{e)}-\textbf{h)} did not. The parameters of the reconstructions are listed in Table \ref{table:parameters} in Appendix \ref{appendix: D}.}
  \label{fig:constraints_length}
\end{figure}

In order to prevent it, we introduced the probe constraint described in Equation \ref{eq: probe_constraint} and took the edges of the predicted patterns into account. The usability of the constants for the parts of bright and dark fields not included into the new basis can be seen in Figure \ref{fig:constraints_length}. Here we used first 45 Zernike polynomials for the description of the bright field and first 22 angular symmetric, i.e. with $m=0$, Zernike dark field polynomials.

\newpage
\section{PCA basis and $l^1$ regularization.}\label{appendix: C}
The concept of a basis generation with the Gram-Schmidt algorithm from patterns recorded before the main acquisition is closely related to principal component analysis (PCA) \cite{PCA_1, PCA_2, PCA_3}. The main difference between our approach and the PCA is that the  Gram-Schmidt compression proposed by us does not require to store the whole dataset. 

\begin{figure}[H]
  \centering
  \includegraphics[width=\textwidth]{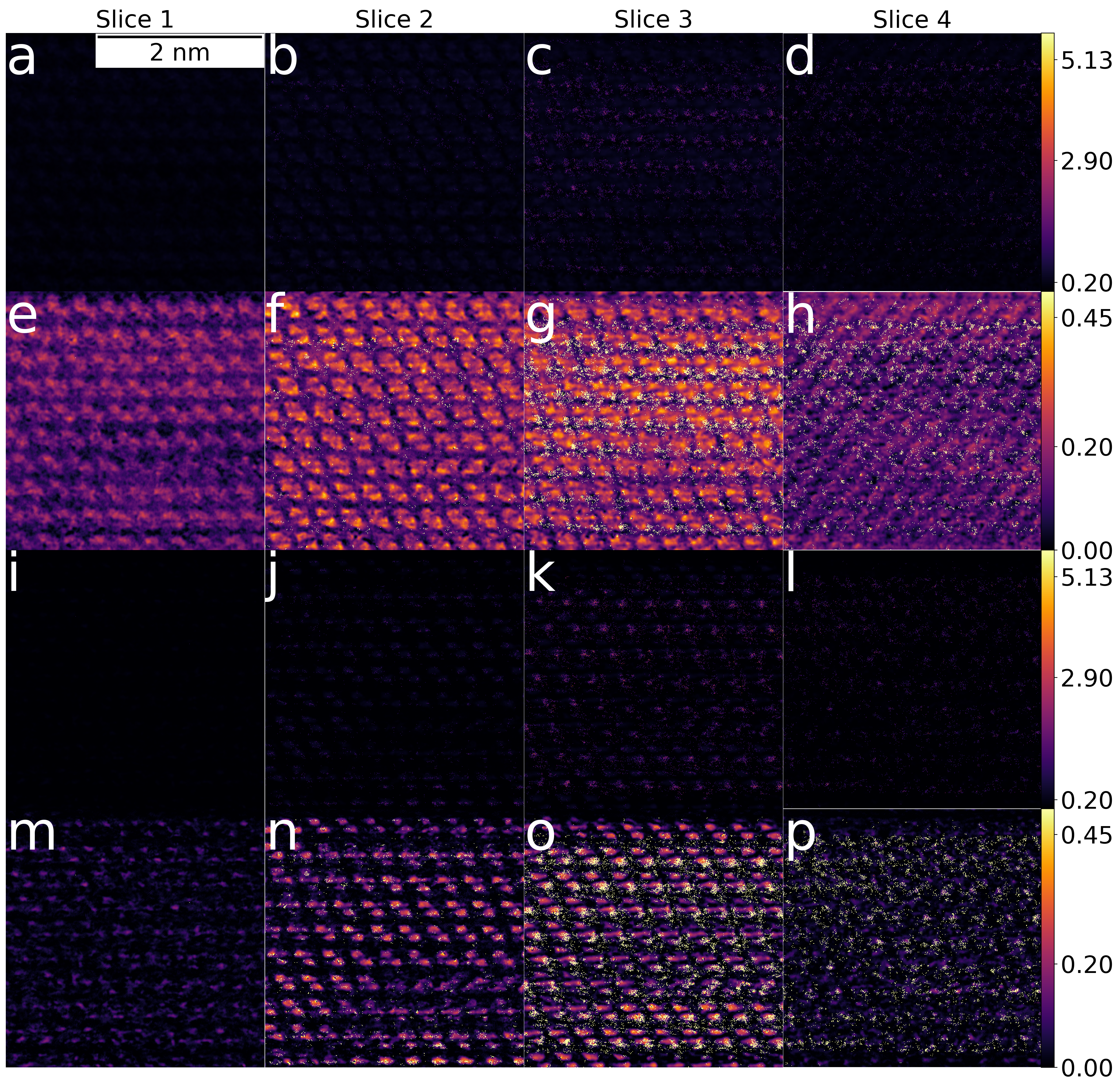}
  \caption{Four multi-slice reconstructions of the phase of object transmission function from a compressed 4D-STEM dataset using 15 PCA components of 320 randomly selected diffraction patterns as a basis. \textbf{a)}-\textbf{d)} and \textbf{e)}-\textbf{f)} show the results of the same reconstruction containing "peak"-formed artifacts, in first row we display the reconstruction with its full dynamic range, and in the second row we set the maximum displayed value to 0.5. \textbf{i)}-\textbf{l)} and \textbf{m)}-\textbf{p)}  show another reconstruction, that was done with the same set of masks, but the weight corresponding to the regularization constraints, total variation of the reconstructed potential and the $l^1$ norm of its phase, were increased by a factor of $10^4$ and $10^7$, respectively. The exact parameters of the two reconstructions are listed in Table \ref{table:parameters} in Appendix \ref{appendix: D}.
  }
  \label{fig:pca_basis}
\end{figure}

To compare the two basis sets we randomly selected 320 diffraction patterns from the initial dataset and generated a basis from first 15 PCA components of the small 4D-STEM dataset. The result of the reconstruction presented first row of Figure \ref{fig:pca_basis} contained artifacts -- point-shaped peaks, which shifted the maximum of the phase up to 6 radians. In the second row of the same Figure we show the same reconstruction, but with an adjusted threshold. As one can see, the reconstructed phase looks correct, but the resolution of the reconstruction is not sufficient to resolve the individual atoms. We obtained the same artifact with "Gram-Schmidt" compression method, but there we could suppress this behaviour by adjusting the $\mathscr{L}_{\perp}$-constraints. Setting higher weights for the $\mathscr{L}_{\perp}$-constraints did not help to prevent the high phase spikes for the PCA basis. We also tried to increase the weight corresponding to a regularization constraints (TV of the object function and $l^{1}$-norm of the reconstructed phase), but, as demonstrated in panels \textbf{i)}-\textbf{p)} of Figure \ref{fig:pca_basis}, this reduced the noise between the atoms, but did not help prevent the appearance of point-shaped peaks. In summary, the basis created through the Gram-Schmidt approach is more robust than the one created via PCA. In addition, the Gram-Schmidt compression is computationally cheaper than the principal component analysis.

\section{Reconstruction parameters}\label{appendix: D}

For the sake of reproducibility of the presented results, we provide the reconstruction parameters. In the following table one can find the learning rates (initial step sizes for the updates of the object function, probe function and the probe positions) and the weights corresponding to the individual parts of the loss function (main part describing the discrepancy between the measurement and algorithms predictions, parts describing the discrepancy between the predictions and the parts of the measured intensity not included into the basis predicted values on the edge of the detector, weight corresponding to the probe constraint and two regularization terms- $l^1$ norm of the reconstructed object and its total variation).

\begin{table}[ht]
\caption{Parameters of the presented reconstruction results.}
    \centering
    \resizebox{\columnwidth}{!}{
    \begin{tabular}{c|c|c|c|c|c|c|c|c|c|c|c}
    \hline
    \multicolumn{1}{c|}{figure}&\multicolumn{1}{c|}{Slice positions in \AA}&\multicolumn{3}{c|}{Learning rates}&\multicolumn{7}{c}{Weights corresponding to the individual parts of the loss function}\\
    
     &  & object  & probe & probe position & ${\mathscr{L}_{main}}$ & ${\mathscr{L}_{\perp}^{bright}}$ & ${\mathscr{L}_{\perp}^{dark}}$ & ${\mathscr{L}_{edge}}$ & ${\mathscr{L}_{Probe}}$  & l1-norm of the phase of the object & TV of the object \\
        \hline
    
    \multicolumn{12}{c}{{Loss function based on uncompressed magnitude}}  \\
        \hline
    \ref{fig:orig}.\textbf{c)}-\textbf{f)} &   [$0$ , $30.7$ ,  $61.4$ , $92$  ] & 1e-3 & 1e-3 & --- &4e-6&--- & ---&--- &---&7e-7&7e-6\\
    \hline
    \multicolumn{12}{c}{{Loss function based on compressed magnitude}}  \\
    \hline
    \ref{fig:dark_field}.\textbf{ a)}-\textbf{d)}& [$0$ , $30.7$ ,  $61.4$ , $92$  ] & 1e-5 & 1e-3 & --- & 2.2e-2&    5e-4 &5e-5 & 5e-2&  5e-2&7e-7&7e-6\\
     \ref{fig:dark_field}.\textbf{ e)}-\textbf{h)}& [$0$ , $30.7$ ,  $61.4$ , $92$  ] & 1e-5 & 1e-3 & --- & 1.5e-2&1e-4 &1e-3 & 5e-2&5e-2&7e-7&7e-6\\
    \ref{fig:dark_field}.\textbf{ i)}-\textbf{l)}& [$0$ , $30.7$ ,  $61.4$ , $92$  ] & 1e-5 & 1e-3 &--- & 2.2e-2& 1e-4 &1e-3 &5e-2 &5e-2&7e-7&7e-6\\
    \ref{fig:dark_field}.\textbf{ m)}-\textbf{p)}& [$0$ , $30.7$ ,  $61.4$ , $92$  ] & 1e-5 & 1e-3 & --- &4.6e-3& 1e-4 &1e-3 & 5e-2&1e-4&7e-7&7e-6\\
    
    \ref{fig:result_comp}.\textbf{a)-d)}&[$0$ , $30.7$ ,  $61.4$ , $92$  ] & 1e-5 & 1e-3 & --- & 2.2e-2& 1e-4& 1e-3&5e-2 & 1e-4&7e-7&7e-6\\
    \ref{fig:result_comp}.\textbf{e)}&[$0$ , $30.7$ ,  $61.4$ , $92$  ] & 1e-5 & 1e-3 & --- & 1.9e-2& 1e-4& 1e-3&5e-2 & 1e-4&7e-7&7e-6\\

    \ref{fig:result_comp}.\textbf{g)-j)}&[$0$ , $30.7$ ,  $61.4$ , $92$  ] & 1e-5 & 1e-3 & --- & 2.7e-2& 5e-5& 5e-4& 5e-2 &5e-2&7e-7&7e-6\\
    \ref{fig:result_comp}.\textbf{k)}&[$0$ , $30.7$ ,  $61.4$ , $92$  ] & 1e-5 & 1e-3 & --- &2.2e-2&  5e-5& 5e-4& 5e-2 &5e-2&7e-7&7e-6\\

    \ref{fig:result_comp}.\textbf{m)-p)}&[$0$ , $30.7$ ,  $61.4$ , $92$  ] & 1e-5 & 1e-3 & --- &6.3e-2& 5e-5& 5e-4& 5e-2 &5e-2&7e-7&7e-6\\
    \ref{fig:result_comp}.\textbf{q)}&[$0$ , $30.7$ ,  $61.4$ , $92$  ] & 1e-5 & 1e-3 & --- & 3.2e-2&  5e-5& 5e-4& 5e-2 &5e-2&7e-7&7e-6\\

    \ref{fig:constraints_length}.\textbf{a)-d)}&[$0$ , $30.7$ ,  $61.4$ , $92$  ] & 1e-5 & 1e-3 & --- & 1.5e-2& 0& 0& 5e-2& 5e-2&7e-7&7e-6\\
    \ref{fig:constraints_length}.\textbf{a)-d)}&[$0$ , $30.7$ ,  $61.4$ , $92$  ] & 1e-5 & 1e-3 & --- &1.5e-2& 1e-4& 1e-3&5e-2 & 5e-2&7e-7&7e-6\\

    \ref{fig:pca_basis}.\textbf{a)-h)}&[$0$ , $30.7$ ,  $61.4$ , $92$  ] & 1e-5 & 1e-3 & --- &6.3e-2&5e-2 &5e-2 & 5e-1& 1e-1&7e-7&7e-6\\
    
    \ref{fig:pca_basis}.\textbf{i)-p)}&[$0$ , $30.7$ ,  $61.4$ , $92$  ] & 1e-5 & 1e-3 & --- &6.3e-2& 5e-2 &5e-2 & 5e-1& 1e-1&7e0&7e-2\\
\hline
    \multicolumn{12}{c}{{Loss function based on compressed intensity}}  \\
             \hline
    \ref{fig:exp_results}.\textbf{a)}& single slice & 1e-3&1e-3& 1e-2 &6.3e-2& 0 & 0 & 0&1e-6 &3e-6&3e-5\\
    \ref{fig:exp_results}.\textbf{b)}& single slice & 1e-3&1e-3& 1e-2 &2.7e-2& 0 & 0 & 0&1e-6&3e-6&3e-5 \\
    \ref{fig:exp_results}.\textbf{c)}& single slice & 1e-3&1e-3& 1e-2 & 2.2e-2& 0 & 0 & 0&1e-6&3e-6&3e-5 \\  
         \hline
    \end{tabular}
    }
    \label{table:parameters}
\end{table}

\end{document}